\documentstyle [twocolumn,prl,aps,floats,epsfig] {revtex}

\begin{document}
\draft
\begin{title}
{Two-dimensional few electron systems in high magnetic fields:\\
Composite-fermion or rotating-electron-molecule approach?}
\end{title} 
\author{Constantine Yannouleas and Uzi Landman} 
\address{
School of Physics, Georgia Institute of Technology,
Atlanta, Georgia 30332-0430 }
\date{February 2002}
\maketitle
\begin{abstract}
A new class of analytic and parameter-free, strongly correlated 
wave functions of simple functional form is derived for few electrons
in two-dimensional quantum dots under high magnetic fields. These wave
functions are constructed through breaking and subsequent restoration
of the circular symmetry, and they offer a natural alternative to the 
Laughlin and composite-fermion functions. Underlying our approach is
a collectively-rotating-electron-molecule picture. The
angular momenta allowed by molecular symmetry correspond to the
filling-factors' hierarchy of the fractional quantum Hall effect. 
\end{abstract}
\narrowtext

Two-dimensional (2D) $N$-electron systems in strong magnetic fields have been 
the focus of extensive theoretical investigations in the last twenty years
\cite{lau1,lau2,jai2,jai1,gir,haw,mac,rua,sek,mak,koo,yl1,yl4}.
The reasons are twofold: (I) The early realization \cite{lau1,lau2} that few
electron systems are relevant to the physics of the fractional quantum Hall
effect (FQHE) observed in the infinite 2D electron gas, and (II) The recent 
progress in nanofabrication techniques that has allowed experiments on 2D 
circular quantum dots (QD's) containing a finite number of electrons 
\cite{ash,kou}.

Among the many theoretical methods for studying such systems, two approaches
have become well established, i.e., exact diagonalization techniques 
\cite{lau1,jai1,rua,sek,mak}  
and consideration of appropriate classes of strongly correlated, analytic 
trial wave functions in the complex plane \cite{lau2,jai2,jai1}. The trial 
wave functions proposed todate have been based on physical intuition, and 
their justification has been inferred {\it a posteriori\/} through comparisons 
with exact numerical calculations and/or with the phenomenology of the FQHE. 

In this paper, we use a systematic, {\it microscopic\/} approach and 
derive a {\it new class\/} of strongly correlated, analytic wave 
functions for the $N$-electron problem in strong magnetic fields \cite{note6}.
Our analytic wave functions have a simple functional form which differs 
from that of the familiar composite-fermion (CF) \cite{jai2} and 
Jastrow-Laughlin (JL) \cite{lau2} functions, and they are associated
with a physical picture of a collectively rotating electron molecule (REM). 
Guiding the synthesis of the states of the system, our approach consists
of two steps: First the breaking of the rotational symmetry at the
single-determinantal {\it unrestricted\/} Hartree-Fock (UHF) level yields
states representing electron molecules (EM's, or finite crystallites).
Subsequently the rotation of the electron molecule is described through 
restoration of the circular symmetry via post Hartree-Fock methods, and in
particular Projection Techniques (PT's) \cite{rs}. Naturally, the restoration 
of symmetry goes beyond the mean-field and yields multi-determinantal wave 
functions. Earlier we demonstrated that this method 
(generalized to include in addition the breaking of the total-spin symmetry)
can describe accurately two-electron systems in molecular \cite{yl2} and
single \cite{yl3} QD's at zero magnetic field \cite{note1}. 

In general, the symmetry-broken UHF \cite{note3} orbitals need to be 
determined numerically \cite{yl1,yl4,yl2,yl3,yl5}. 
However, in the case of an infinite 2D electron gas in strong magnetic fields,
it has been found \cite {mz} that such UHF
orbitals \cite{yfl} can be approximated by analytic Gaussian functions 
centered at different positions $Z_j \equiv X_j+ \imath Y_j$ and forming an 
hexagonal Wigner crystal (each Gaussian representing a localized electron). 
The specific expression for these displaced Gaussians is
\begin{eqnarray}
u(z, &&Z_j) = (1/\sqrt{\pi}) \nonumber \\
&& \times \exp[-|z-Z_j|^2/2] \exp[-\imath (xY_j+yX_j)]~,
\label{gaus}
\end{eqnarray}
where the phase factor is due to the gauge invariance. $z \equiv x-\imath y$,
and all lengths are in dimensionless units of ${l_B}\sqrt{2}$ with 
the magnetic length being $l_B=\sqrt{\hbar c/eB}$. 

In the case of a Coulombic finite $N$-electron system, it has been found 
\cite{koo,yl1} that the UHF orbitals arrange in concentric rings forming EM's 
(referred to also as Wigner molecules, WM's) \cite{note2}. The UHF results for
the formation of WM's are in agreement with the molecular structures obtained 
via the conditional probability distributions (CPD's) which can be extracted 
from exact numerical wave functions \cite{sek,mak,yl6}. For  $N \leq 4$, the 
electrons are located at the apexes of a regular polygon situated on a single
ring, while for $ 5 \leq N \leq 7$ both the single-ring structure and an 
isomeric one with one electron at the center come into play. We will denote 
the former arrangement as $(0,N)$ and the latter as $(1,N-1)$. The electrons 
of the $(0,N)$ ring are located at \begin{equation}
Z_j=Z\exp[\imath 2\pi (1-j)/N],\;\;  1 \leq j \leq N~,
\label{zj1}
\end{equation}
and those participating in a $(1,N-1)$ arrangement are located at 
\begin{equation}
Z_1=0;\; Z_j=Z \exp[\imath 2\pi (2-j)/(N-1)], \;\; 2 \leq j \leq N~.
\label{zj2}
\end{equation}

Before proceeding further, we need to expand the displaced Gaussian 
(\ref{gaus}) over the Darwin-Fock single-particle states. Due to the high 
magnetic field, only the single-particle states,
\begin{equation}
\psi_l(z) = \frac{z^l}{\sqrt{\pi l!}} \exp(-zz^*/2)~,
\label{lll}
\end{equation}
of the lowest Landau level (LLL) are needed (observe that the angular
momentum of this state is $-l$ due to the definition $z \equiv x-\imath y$).
Then a straightforward calculation \cite{mik} yields
\begin{equation}
u(z,Z)=\sum_{l=0}^{\infty} C_l(Z) \psi_l(z)~,
\label{uexp}
\end{equation}
with $C_l(Z)=(Z^*)^l \exp(-ZZ^*/2)/\sqrt{l!}$ for $Z \neq 0$. Naturally,
$C_0(0)=1$ and $C_{l>0}(0)=0$.

Since electrons in strong magnetic fields are fully polarized,
only the space part of the many-body wave functions needs to be considered;
for the symmetry-broken UHF determinant describing the WM, it is given by
\begin{equation}
\Psi_{\text{UHF}}^N = {\text{det}} [u(z_1,Z_1),u(z_2,Z_2),\cdot \cdot \cdot,
u(z_N,Z_N)]~.
\label{uhfd}
\end{equation}
Using (\ref{uexp}) one finds the following expansion (within a 
proportionality constant)
\begin{eqnarray}
\Psi_{\text{UHF}}^N =&& \sum_{l_1=0,...,l_N=0}^{\infty} 
\frac{ C_{l_1}(Z_1)C_{l_2}(Z_2) \cdot \cdot \cdot C_{l_N}(Z_N) }
{\sqrt{l_1! l_2! \cdot \cdot \cdot l_N!} }  \nonumber \\
&& \times \; D(l_1,l_2,...,l_N) \exp(-\sum_{i=1}^N z_i z_i^*/2)~,
\label{uhfde}
\end{eqnarray}
where $D(l_1,l_2,...,l_N) 
\equiv {\text{det}}[z_1^{l_1},z_2^{l_2}, \cdot \cdot \cdot, z_N^{l_N}]$. 

The UHF determinant [Eq.\ (\ref{uhfd}) or Eq.\ (\ref{uhfde})] breaks the
rotational symmetry and thus it is is not an eigenstate of the total angular 
momentum $\hbar \hat{L}=\hbar \sum_{i=1}^N \hat{l}_i$. However, one can 
{\it restore\/} the rotational symmetry by applying onto the UHF determinant 
the following projection operator \cite{yl3,rs}
\begin{equation}
2 \pi {\cal P}_L \equiv \int_0^{2 \pi}
d\gamma \exp[i \gamma (\hat{L}-L)]~,
\label{amp}
\end{equation}
where $\hbar L=\hbar \sum_{i=1}^N l_i$ are the eigenvalues of the total 
angular momentum. 

It is advantageous to operate with ${\cal P}_L$ on expression (\ref{uhfde}),
which is an expansion in a basis consisting of products of single-particle
eigenstates with good angular momenta $l_i$. Indeed in this case the 
projection operator acts as a Kronecker delta: from the unrestricted sum
(\ref{uhfde}), it picks up only those terms having a given total angular
momentum $L$. As a result, after taking into consideration the specific
electron locations (\ref{zj1}) associated with the $(0,N)$ WM, one derives 
\cite{note4} the following symmetry-preserving, many-body correlated wave 
functions (within a proportionality constant),
\begin{eqnarray}
\Phi_L^N && =
 \sum^{l_1 + \cdot \cdot \cdot +l_N=L}%
_{0 \leq l_1<l_2< \cdot \cdot \cdot <l_N}
\left( \prod_{i=1}^N l_i! \right)^{-1}  \nonumber \\
&& \times \left( \prod_{1 \leq i < j \leq N} 
\sin \left[\frac{\pi}{N}(l_i-l_j)\right] \right)  \nonumber \\
&& \times \; D(l_1,l_2,...,l_N)
\exp(-\sum_{i=1}^N z_i z_i^*/2)~.
\label{phi1}
\end{eqnarray}
In deriving (\ref{phi1}), we took into account that for each determinant 
$D(l_1,l_2,...,l_N)$ in the unrestricted expansion (\ref{uhfde})
there are $N!-1$ other determinants generated from it 
through a permutation of the indices $\{l_1,l_2,...,l_N\}$; these determinants
are equal to the original one or differ from it by a sign only.
In the case of an $(1,N-1)$ WM, the corresponding correlated
wave functions are given by,
\begin{eqnarray}
\Phi^{\prime N}_L && = 
\sum^{l_2+ \cdot \cdot \cdot +l_N=L}%
_{0 \leq l_2 < l_3 < \cdot\cdot\cdot < l_N}
\left( \prod_{i=2}^N l_i! \right)^{-1} \nonumber \\
&& \times \left( \prod_{2 \leq i < j \leq N} 
\sin \left[\frac{\pi}{N-1}(l_i-l_j)\right] \right)  \nonumber \\
 \nonumber \\
&& \times \; D(0,l_2,...,l_N)
\exp(-\sum_{i=1}^N z_i z_i^*/2)~.
\label{phi2}
\end{eqnarray}

We call the correlated wave functions [Eq.\ (\ref{phi1}) 
and Eq.\ (\ref{phi2})] the electron-molecule wave functions (EMWF's).
We stress that the EMWF's have good total angular momenta, unlike the 
UHF determinant from which they were projected out. The 
projection operator (\ref{amp}) acts on a single UHF determinant, but yields 
a whole rotational band of the WM. The states in this band are those with
the lowest energy for a given angular momentum $L$, and in addition they are 
{\it purely\/} rotational, i.e., they carry no other internal excitations; 
in analogy with the customary terminology from the spectroscopy of rotating
nuclei \cite{yl6,bm}, we designate this band as the ``yrast band''.

Furthermore, if instead of electrons the displaced Gaussians 
(\ref{gaus}) describe {\it bosonic\/} particles forming a molecule, 
the corresponding \cite{yl3} many-body correlated wave functions will be given
by expressions similar to Eq.\ (\ref{phi1}) and Eq.\ (\ref{phi2}), but with
the following two important differences: (I) The product of sine functions 
will be replaced by a sum over cosines, and (II) The 
determinants $D(l_1,l_2,...,l_N)$ will be replaced by permanents \cite{note5}
$P(l_1,l_2,...,l_N) \equiv {\text{perm}} [z_1^{l_1},z_2^{l_2},...,z_N^{l_N}]$.

Among the properties of the EMWF's specified by Eq.\ (\ref{phi1}) and 
Eq.\ (\ref{phi2}), we mention the following:

1) The EMWF's lie entirely within the Hilbert subspace spanned by the lowest 
Landau level and, after expanding the determinants \cite{note4}, they can be 
written in the form (within a proportionality constant),
\begin{equation}
\Phi^N_L[z] = P^N_L [z] \exp(-\sum_{i=1}^N z_i z^*_i/2)~,
\label{pl}
\end{equation}
where the $P^N_L[z]$'s are order-$L$ homogeneous polynomials of the $z_i$'s.

2) The polynomials $P^N_L[z]$ are divisible by 
\begin{equation}
P^N_V[z] = \prod_{1 \leq i < j \leq N} (z_i-z_j)~,
\label{pv}
\end{equation}
namely $P^N_L[z]=P^N_V[z] Q^N_L[z]$.
This is a consequence of the antisymmetry of $\Phi^N_L[z]$.
$P^N_V[z]$ is the Vandermonde determinant $D(0,1,...,N)$. For the case of
the lowest allowed angular momentum $L_0=N(N-1)/2$ (see below), one
has $P^N_{L_0}[z] = P^N_V[z]$, a property that is shared with the 
Jastrow-Laughlin \cite{lau2} and composite-fermion \cite{jai2} trial 
wave functions.

3) Upon the introduction of the Jacobi coordinates, the center-of-mass 
separates from the internal variables in complete analogy with the exact
solution.

4) The coefficients of the determinants [i.e., products of sine functions, see
Eq.\ (\ref{phi1} and Eq.\ (\ref{phi2})] dictate that the EMWF's are nonzero 
only for special values of the total angular momentum $L$ given by,
\begin{equation}
L=N(N-1)/2 + N k,\;\;k=0,1,2,3,...~,
\label{l0n}
\end{equation}
for the $(0,N)$ configuration, and
\begin{equation}
L=N(N-1)/2 + (N-1) k,\;\;k=0,1,2,3,...~,
\label{l1nm1}
\end{equation}
for the $(1,N-1)$ one. The minimum angular momentum $L_0=N(N-1)/2$ is 
determined by the fact that the $D$ determinants [see Eq.\ (\ref{phi1}) and
Eq.\ (\ref{phi2})] vanish if any two of the single-particle angular momenta 
$l_i$ and $l_j$ are equal. 
In plots of the energy vs. the angular momenta, derived from 
exact-diagonalization studies 
\cite{gir,haw,mac,rua,sek,mak}, it has
been found that the special $L$ values given by Eq.\ (\ref{l0n}) and 
Eq.\ (\ref{l1nm1}) exhibit prominent cusps
reflecting enhanced stability; as a result these $L$ values are 
often referred to as ``magic angular momenta''. We stress that 
the angular momenta associated with the EMWF's correspond precisely to the 
magic $L$'s of the exact-diagonalization studies \cite{note12}.
In the thermodynamic limit \cite{lau2,gir}, one 
can relate the total $L$ to a fractional filling through the
relation  $\nu=N(N-1)/(2L)$,
and thus the EMWF angular momenta (\ref{l0n}) and (\ref{l1nm1})
correspond to all the fractional filling factors associated with the FQHE,
including the even-denominator ones, i.e., $\nu=$ 1, 3/5, 3/7, 5/7, 2/3,
1/2, 1/3, etc... 

\begin{table}[t]
\caption{The $Q^3_9[z]$ polynomial associated with the EMWF's and the JL 
functions (The $Q^N_L[z]$ polynomials are of order $L-L_0$).}
\begin{tabular}{cc}
EMWF & $~~(z_1^3 -3 z_1^2 z_2 + z_2^3 +6 z_1 z_2 z_3 -3 z_2^2 z_3 -3 z_1 z_3^2
        +z_3^3) $ \\
~& $\times (z_1^3 -3 z_1 z_2^2 + z_2^3 +6 z_1 z_2 z_3 -3 z_1^2 z_3 -3 z_2 z_3^2
        +z_3^3)$ \\ \tableline
JL & $(z_1-z_2)^2 (z_1-z_3)^2 (z_2-z_3)^2$ \\
\end{tabular}
\end{table}

5) For the case of two electrons $(N=2)$, the EMWF's reduce to
the Jastrow-Laughlin form, namely
\begin{equation}
P^2_L[z]=\prod_{1 \leq i < j \leq N} (z_i-z_j)^L~,
\label{p2}
\end{equation}
where $L=1$, 3, 5, ...
However, this is the only case for which there is coincidence between 
the EMWF's and the JL wave functions. 
For higher numbers of electrons, $N$, the EMWF polynomials $P^N_L[z]$ 
(apart from the lowest-order Vandermonde $P^N_{L_0}[z]$ ones) are quite
different from the corresponding JL or composite-fermion polynomials.
In particular, the familiar factor $\prod_{1 \leq i < j \leq N} 
(z_i-z_j)^{2p}$, with $p$ an integer \cite{jai2,jai1}, 
(which reflects multiple zeroes) does not appear in the EMWF's (see, e.g., 
Table I which contrasts the $Q^3_9[z]$ polynomials corresponding to the 
EMWF's and JL functions).

6) For the case of three electrons $(N=3)$, after transforming to the
Jacobi coordinates 
$\bar{z}=(z_1+z_2+z_3)/3$, $z_a=(2/3)^{1/2}((z_1+z_2)/2-z_3)$,
$z_b=(z_1-z_2)/\sqrt{2}$ (and dropping the center-of-mass exponential
factor), the EMWF's can be written as (again within a proportionality
constant),
\begin{eqnarray}
\Phi^3_L[z_a,z_b]=&& [(z_a+ \imath z_b)^{L}-(z_a-\imath z_b)^{L}]
\nonumber \\
&& \times \exp[(-1/2)(z_a z_a^* + z_b z_b^*)]~,
\label{jac}
\end{eqnarray}
with $L=3m$, $m=1$, 2, 3, 4, ... being the total angular momentum. 
Again the wave functions $\Phi^3_L[z_a,z_b]$ are very different from the 
three-electron JL ones; e.g., they are nonvanishing for even $m$
values, unlike the three-electron JL functions. However, the
$\Phi^3_L[z_a,z_b]$'s coincide with the functions $|m,0 \rangle$ derived 
in Ref.\ \cite{lau1}. We notice that, although it was found \cite{lau1,lau3}
that these wave functions exhibited behavior expected of fractional
quantum Hall ground states, the generalization of them to a higher
number of electrons did not follow.

\begin{table}[t]
\caption{Overlaps, $\langle \phi^N_L|\psi^N_L \rangle / 
(\langle \phi^N_L|\phi^N_L \rangle \langle \psi^N_L|\psi^N_L \rangle)^{1/2}$, 
of EMWF's ($\phi$'s) and JL functions ($\phi$'s) with the corresponding exact 
eigenstates ($\psi$'s) for various values of the angular momenta $L$.
Recall that the angular momenta for the JL functions are $L_{JL}=N(N-1)m/2$,
with $m > 0$ being an odd integer. Bottom: Energies of EMWF's compared to 
CF and exact-diagonalization results. Energies in units of $e^2/\kappa l_B$,
($\kappa$ is the dielectric constant).}
\begin{tabular}{crlll}
  OVERLAPS    &  $L$ & EMWF     & JL  &     ~ \\ \tableline
 N=3    & 9   & 0.98347  & 0.99946\tablenotemark[1] & ~\\
 ~     & 15   & 0.99473  & 0.99468\tablenotemark[1] & ~\\
 ~     & 21   & 0.99674  & 0.99476\tablenotemark[1] & ~\\
 ~     & 27   & 0.99758  & 0.99573\tablenotemark[1] & ~\\
 ~     & 33   & 0.99807  & 0.99652\tablenotemark[1] & ~\\
 ~     & 39   & 0.99839  & 0.99708\tablenotemark[1] & ~ \\ \tableline
 N=4    & 18   & 0.92937  & 0.97880 & ~\\
 ~     & 30   & 0.96742  & 0.94749 & ~\\
 ~     & 42   & 0.97366  & 0.95561 & ~\\
 ~     & 54   & 0.97623  & 0.96815 & ~\\ \tableline
 ENERGIES    &  $L$ & EMWF     & CF  & EXACT   \\ \tableline 
 N=4  &  10  & 1.78510  & 1.78537\tablenotemark[2]  & 1.78509 \\
 ~    &  14  & 1.50955  & 1.50222\tablenotemark[2]  & 1.50066 \\
\end{tabular}
\tablenotetext[1] {From Ref.\ \cite{lau2}.}
\tablenotetext[2] {From Table V of Ref.\ \cite{jai3}.}
\end{table}


Several publications \cite{jai1,haw,jai3} have applied the composite fermion
picture (the JL functions are a special case of the CF's)
to single QD's in strong magnetic fields. In particular, it has been shown
\cite{jai1} that CF wave functions can be constructed with
angular momenta coinciding with the magic ones. However, it has also been 
found \cite{sek} that several discrepancies exist, i.e.,
some of the larger magic angular momenta are not reproduced by the CF picture.
As a consequence of the above, the REM description with the EMWF's derived 
here offers a natural alternative for interpreting the physics of electrons
in QD's in high magnetic fields. This proposition is further supported
by inspection of the overlaps between the EMWF's and the exact many-body
eigenstates, and their comparison with the corresponding overlaps for the
JL states; see Table II, where in some instances (i.e., $N=4$, $L=10$ and
14) we list energies of the EM, CF, and exact states instead of the
overlaps. Indeed the agreement between the EM states and the exact ones
is of comparable quality as in the case of the CF and JL wave functions. 

In summary, we have developed a new class of analytic and parameter-free,
strongly correlated wave functions of simple functional form, which accurately
describe the physics of electrons in QD's under high magnetic fields. The 
thematic basis of our approach is built upon the intuitive picture of 
collectively rotating electron molecules, and the synthesis of the many-body 
EMWF's involves breaking of the circular symmetry at the UHF level with 
subsequent restoration of this symmetry via a projection technique. While we 
focus here on the strong magnetic-field regime, we note that the REM picture 
unifies the treatment of strongly correlated states of electrons in QD's over 
the whole magnetic-field range \cite{yl1,yl5,yl6}. We also remark
that our analysis, aimed here mainly at treating finite electron systems
(i.e., QD's) with an arbitrary number of electrons, points to the remarkable 
conclusion that the observed FQHE hierarchy of filling factors may be viewed 
as an experimental signature of the yrast band (see above) of the REM.

This research is supported by the U.S. D.O.E. (Grant No. FG05-86ER-45234).


\begin{references}
\bibitem{note6}
Since we are interested in obtaining analytic results, we specifically
consider the limit when any external potential confinement can be 
neglected compared to the confinement induced by the magnetic field.
\bibitem{note3}
In our earlier publications, we have used the term
spin-and-space (sS)-UHF to emphasize the breaking of both 
the spin and space symmetries. In high magnetic fields,
the electrons are fully polarized and only breaking of
the space symmetry is involved (S-UHF). For convenience, 
in this paper we have dropped the prefix S when referring to the 
S-UHF.
\bibitem{note2}
Depending on the strength of the interelectron repulsion,
WM's can also form at zero magnetic field, see Refs.\ \cite{yl1,yl5}.
For a CPD/exact-solution study of such WM's at $B=0$, see Ref.\ \cite{yl6}.
\bibitem{note4}
The algebraic manipulations reported here were performed using
MATHEMATICA.
\bibitem{note5}
An analog of a determinant where all the signs in the expansion by minors are
taken as positive.
\bibitem{note12}
Magic angular momenta have been also discussed in the context of
the geometric configuration \cite{rua,sek} and the Eckart frame
\cite{mak} models. However, unlike our approach, these models do not
afford derivation of analytical forms for the EMWF's.
\end{references}
\end{document}